# Giant Hole-doping in 2*H*-WSe$_2$ via Ta Substitution

Minhee Kang[a,†], Woojin Choi[b,†], Choongyu Hwang[a] and Jinwoong Hwang[b,*]

[a]Department of Physics, Pusan National University, Busan 46241, Republic of Korea

[b]Department of Semiconductor physics and Institute of Quantum Convergence Technology, Kangwon National University, Chuncheon 24341, Republic of Korea

**Abstract**

The family of transition metal dichalcogenides (TMDCs) has been regarded as promising candidates for future electronics, valleytronics, spintronics, and optoelectronics. While most of TMDCs are intrinsic *n*-type semiconductors due to electron donation from chalcogen vacancies, realizing intrinsic *p*-type TMDCs and achieving precise control over their electronic properties remain challenging. In this work, we introduce a powerful approach to obtain intrinsic hole doping by substituting Ta atom into 2*H*-WSe$_2$. A combining study of molecular beam epitaxy growth and *in-situ* angle-resolved photoemission spectroscopy characterization clearly reveals that Ta substitution induces a significant hole doping and provides a possible way of a semiconductor-to-metal transition in 2*H*-WSe$_2$.

† These authors contributed equally to this work

* Corresponding author

Email: jwhwang@kangwon.ac.kr

# 1. Introduction

Transition metal dichalcogenides (TMDCs) are a class of van der Waals (vdW) layered materials characterized by a sandwich-like structure, in which hexagonally packed transition metal atoms are sandwiched between two layers of chalcogen atoms. The combination of different transition metals and chalcogens not only determines their intrinsic electronic structures but also gives rise to a variety of emergent quantum phases, including superconductivity, charge density waves, and magnetism [1-3]. Furthermore, the weak interlayer vdW interaction allows TMDCs to be mechanically or epitaxially thinned down to the monolayer limit, enabling the exploration of distinct physical phenomena emerging in reduced dimensions [2,3]. By varying the thickness of TMDC layers, one can tune quantum confinement, dielectric screening, and interlayer coupling, thereby realizing novel ground states that do not exist in bulk counterparts [1-3]. Owing to these unique features, TMDCs have become one of the central platforms in two-dimensional material research, offering vast opportunities to explore fundamental physics and develop next-generation electronic and quantum devices [4].

Among the family of TMDCs, 2$H$-WSe$_2$ has attracted considerable attention due to its semiconducting nature with strong light-matter coupling and the largest spin-splitting in the valence band at the K point, making it an ideal platform for the next generation electronic and photonic devices [1,4-8]. While $n$-type 2$H$-WSe$_2$ can be readily obtained and widely controlled through chalcogen vacancies or ionic gating [9-11], realizing $p$-type 2$H$-WSe$_2$ remains challenging even though both $p$- and $n$-type TMDCs are indispensable for complementary device applications. Although engineering extrinsic factors can induce $p$-type behaviors [12,13], it is still difficult to effectively control the hole-doping levels in 2$H$-WSe$_2$. Considering the small effective hole mass (0.45$m_0$) and high hole mobility [14,15], developing intrinsic high-quality $p$-type 2$H$-WSe$_2$ is highly desirable as an excellent platform for exploring novel quantum phenomena and device applications.

In this work, we report a successful synthesis of Ta-substitutional 2$H$-WSe$_2$ film using molecular beam epitaxy (MBE) and a characterization of its electronic structure using angle-resolved photoemission spectroscopy (ARPES). Our experimental results reveal that 2$H$-WSe$_2$ bands move toward Fermi energy ($E_F$) as much as 760 meV by Ta substitution. while the strength of band split at the K point is hardly changed compared to stoichiometric 2$H$-WSe$_2$. We further find clear evidence for the partial existence of a metallic band, crossing $E_F$ in 2$H$-WSe$_2$ by the introduction of Ta atoms. Our results not only provide a powerful strategy to achieve a significant hole-doping and metallization of 2$H$-WSe$_2$ but also pave the way for future electronic and spintronic applications.

## 2. Experimental details

Both 2$H$-WSe$_2$ and Ta-substituted WSe$_2$ films (2$H$-W$_{(1-x)}$Ta$_{(x)}$Se$_2$) were grown by MBE on epitaxially grown bilayer graphene on 6$H$-SiC(0001). High-purity W (99.999%) and Ta (99.9%) were evaporated from an $e$-beam evaporator, and Se (99.999%) was evaporated by a standard Knudsen effusion cell. Few layers of 2$H$-WSe$_2$ films were grown with a flux ratio W:Se = 1:20, and the substrate temperature was held at 400 ˚C during the growth. This yields the growth rate of 25 minutes per monolayer monitored by *in-situ* reflection high-energy electron diffraction (RHEED). After growth, 2$H$-WSe$_2$ film was annealed at 430 ˚C for 1 hour to improve film quality. 2$H$-W$_{(1-x)}$Ta$_{(x)}$Se$_2$ was grown by co-evaporation with Ta and Se on MBE-grown 2$H$-WSe$_2$. The flux ratio was Ta:Se = 1:20 (the absolute Ta flux is same to W flux), and the substrate temperature was held at 450 ˚C during the growth for 30 minutes, yielding a growth rate of 25 minutes per monolayer. *in-situ* RHEED measurement was carried out with a high voltage of 20 kV throughout the growth process. The base pressure of the MBE chamber was 5×10$^{-10}$ Torr. The MBE-grown samples were then transferred under an ultrahigh-vacuum (UHV) environment into the ARPES analysis chamber for the measurement at the HERS end station of Beamline 10.0.1, Advanced Light Source, Lawrence Berkeley National Laboratory. ARPES data were taken using a Scienta R4000 analyzer at a base pressure of 3×10$^{-11}$ Torr. The photon energy was set at 70 eV for $s$-polarization with energy and angular resolutions of 18-25 meV and 0.1˚, respectively. The spot size of the photon beam on the sample was ~100 μm×100 μm. Potential charging effect from the insulating samples has been monitored by checking reference spectra with varying photon flux.

## 3. Results and discussion

Figure 1(a) illustrates the MBE growth process for $2H$-W$_{(1-x)}$Ta$_{(x)}$Se$_2$ films: First, 5 layers of the $2H$-WSe$_2$ film were grown on bilayer graphene on an SiC substrate (the left panel of Fig. 1(a)). Second, $2H$-W$_{(1-x)}$Ta$_{(x)}$Se$_2$ was synthesized by co-evaporation of Ta and Se on the MBE-grown $2H$-WSe$_2$ for Ta-substitution (the right panel of Fig. 1(a)). Figures 1(b) and 1(c) show the RHEED images of MBE-grown $2H$-WSe$_2$ and $2H$-W$_{(1-x)}$Ta$_{(x)}$Se$_2$ film, respectively. Clean vertical line profiles in RHEED images after the growth indicate the well-defined formation of films. A lattice constant of stoichiometric WSe$_2$ is estimated to be ~3.29 Å. After co-evaporations of Ta and Se, the vertical lines become close and broad (the lower panel of Fig. 1(b)) compared to stoichiometric $2H$-WSe$_2$ (the upper panel of Fig. 1(b)). The lattice constant of $2H$-W$_{(1-x)}$Ta$_{(x)}$Se$_2$ extracted from RHEED peaks shows ~3.40 Å which is close to a value of $1H$-TaSe$_2$ (~3.43 Å) [5,16]. The angle-integrated core-level spectrum of $2H$-WSe$_2$ (blue symbols) displays characteristic peaks for W $4f$ and Se $3d$, demonstrating the film's high purity (Figs. 1(c) and 1(d)). On the other hands, the $2H$-W$_{(1-x)}$Ta$_{(x)}$Se$_2$ (red symbols) film shows multiple Se $3d$ and additional Ta $4f$ peaks (Figs. 1(c) and 1(d)). W and Ta $4f$ peaks show characteristic two peaks by spin-orbit splitting [5,17] and Se $3d$ peaks exhibit three peaks. Considering the multiple peaks of Se $3d$, which correspond to the peak positions for $2H$-WSe$_2$ and $2H$-TaSe$_2$ [5,17], and the reduced lattice constant in $2H$-W$_{(1-x)}$Ta$_{(x)}$Se$_2$ film, these results may indicate a possible formation of $2H$-W$_{(1-x)}$Ta$_{(x)}$Se$_2$ as well as $1H$-TaSe$_2$ films on WSe$_2$.

To investigate the origin of the changes observed in the RHEED and the multiple core-level spectra of $2H$-W$_{(1-x)}$Ta$_{(x)}$Se$_2$, we performed *in-situ* ARPES measurements. Figure 2(a) presents an ARPES intensity map of 5 layers of the MBE-grown $2H$-WSe$_2$ film taken along the Γ-K direction at room temperature (RT). The valence band maximum is located at the Γ-point rather than at the K-point, which well corresponds to the multilayer $2H$-WSe$_2$ [5]. The valence bands split into two nearly spin-degenerate bands upon dispersing towards the K point. The split valence bands are more clearly visualized in the second derivative ARPES intensity map as shown in Fig. 2(b). The size of spin split bands at the K-point extracted from the energy distribution curve (EDC) taken at the K-point is ~480 meV (the upper panel of Fig. 2(e)) [5].

Figures 2(c) and 2(d) present the raw and second derivative ARPES intensity maps of $2H$-W$_{(1-x)}$Ta$_{(x)}$Se$_2$ taken along the Γ-K direction at RT, respectively. The introduction of Ta atoms into the MBE-grown $2H$-WSe$_2$ system induces two notable changes in the electron band structure: (i) The entire bands of WSe$_2$ shift towards $E_F$ as much as ~760 meV, extracted from the EDC taken at the K-point as shown in the lower panel of Fig. 2(e). (ii) Unexpected hole bands at the Γ-point and an electron band in the middle of the Γ-K direction show up as denoted by black and red dashed curves as a guideline, respectively, in Fig. 2(d). Considering the shape of band dispersions and the peak positions of the core-level spectra (Figs. 1(c) and 1(d)) [16,17], the emergence of the additional bands in the ARPES spectra of $2H$-W$_{(1-x)}$Ta$_{(x)}$Se$_2$ originates from the partial formation of monolayer $1H$-TaSe$_2$ on $2H$-WSe$_2$ films. Detailed analysis of the electron band structure of $2H$-W$_{(1-x)}$Ta$_{(x)}$Se$_2$ further reveals that the split size of the valence bands at the K-point still remains the same, i.e., ~480 meV (Fig. 2(e)). Since the size of the band splitting at the K-point depends on the strength of the spin-orbit coupling of transition metal atoms [18], the similar atomic mass of Ta to W does not give additional band splitting at the K-point.

It is noteworthy that there are additional peaks with weak spectral intensity in $2H$-W$_{(1-x)}$Ta$_{(x)}$Se$_2$ at the K-point as denoted by the black arrows in Fig. 2(e). To identify this feature, ARPES data for $2H$-WSe$_2$ are compared to those of $2H$-W$_{(1-x)}$Ta$_{(x)}$Se$_2$ in Fig. 3. The raw and second derivative ARPES intensity maps for the stoichiometric $2H$-WSe$_2$ film shown in Figs. 3(a) and 3(b), respectively, clearly present two split bands at the K-point as denoted by blue arrows. On the other hand, ARPES data for the $2H$-W$_{(1-x)}$Ta$_{(x)}$Se$_2$ film shown in Figs. 3(c) and 3(d) present four split bands at the K-point, two bands with strong spectral intensity (blue arrows) and other two with weak intensity (black arrows). One of the latter two crosses $E_F$, forming a Fermi surface (Fig. 3(f)). Considering that $1H$-TaSe$_2$ also exists in the MBE-grown $2H$-W$_{(1-x)}$Ta$_{(x)}$Se$_2$ sample as evidenced in core-level spectra shown in Figs. 1(c) and 1(d), the metallic bands are clearly resolved when the Fermi surfaces of pristine ML $1H$-TaSe$_2$ and $2H$-W$_{(1-x)}$Ta$_{(x)}$Se$_2$ films are compared to each other as shown in Figs. 3(e) and 3(f). Indeed, the Fermi surface of MBE-grown $2H$-W$_{(1-x)}$Ta$_{(x)}$Se$_2$ film shown in panel (f) shows additional features at both Γ and K points as denoted by red dashed circles in Fig. 3(f), which are not observed in pristine ML 1$H$-TaSe$_2$ shown in panel (e). These results indicate that semiconducting 2$H$-WSe$_2$ has a metallic character with a giant hole doping via Ta substitution.

The hole-doped 2$H$-WSe$_2$ by Ta substitution itself is not surprising since the group V transition metals (V, Ta, Nb) have one fewer valence electron than W, offering a direct way to introduce hole carriers [19]. However, the metallic character of 2$H$-W$_{(1-x)}$Ta$_{(x)}$Se$_2$ is not trivial. Although several works have reported $p$-type 2$H$-WSe$_2$ prepared by extrinsical molecular doping and surface treatments [12,13], the conventional methods still give semiconducting nature of 2$H$-WSe$_2$. On the other hands, our work clearly shows metallic nature of 2$H$-WSe$_2$ by Ta substitution, providing the possible way to induce a semiconductor-to-metal transition in 2$H$-WSe$_2$. Moreover, recent studies on Ising superconductivity suggest that the hole-doped metallic 2$H$-WSe$_2$ could be a promising platform to realize the topological Ising superconducting phase due to the strong spin-orbit coupling and broken in-plane inversion symmetry [20,21]. Consequently, 2$H$-W$_{(1-x)}$Ta$_{(x)}$Se$_2$ can be a strong candidate to explore novel spin-valley-coupled superconducting states.

Another unusual feature observed in 2$H$-W$_{(1-x)}$Ta$_{(x)}$Se$_2$ is the split four valence bands at the K point as shown in Figs. 3(c) and 3(d). The additional band splitting can be attributed to the formation of randomly distributed Ta clusters [22], where Ta atoms are irregularly substituted into W sites of 2$H$-WSe$_2$, leading to local variations in the electronic potential [23]. Alternatively, the splitting in 2$H$-W$_{(1-x)}$Ta$_{(x)}$Se$_2$ may also originate from the emergence of local magnetic moments driven by clustering of Ta atoms [23]. It is recently reported that localized magnetic moments could be induced by clustering Ta atoms in 2$H$-WTaSe$_2$ [23]. Ta clusters with a triangular shape, arranged in the ordered manner were observed in STM measurements [23]. Calculations within the density functional theory suggests that the Ta cluster induces local strain, producing local magnetic moments [23]. The ferromagnetic ordering of the local magnetic moments could give rise to energy band splitting and the additional splitting that occurs not only at the K point but also around the Γ point (Fig. 3) could be explained by the possible ferromagnetic moments. Further investigation is required to understand the non-trivial band splitting at the K points of 2$H$-W$_{(1-x)}$Ta$_{(x)}$Se$_2$ with homogeneous bulk samples.

**4. Conclusions**

Electronic properties of MBE-grown Ta-substituted 2$H$-WSe$_2$ films have been investigated using the ARPES technique. The introduction of Ta atoms in the system allows 2$H$-WSe$_2$ to exhibit $p$-type nature and to undergo a semiconductor-to-metal transition with unexpected valence band splitting at the K point. Our findings provide not only an experimental evidence of giant hole doping in 2$H$-WSe$_2$ by Ta substitution, but also a promising route to engineer the electronic properties in two-dimensional van der Waals materials

**Acknowledgements**

This research was supported by 2023 Research Grant from Kangwon National University.

**Conflicts of Interest**

The authors declare no conflicts of interest.

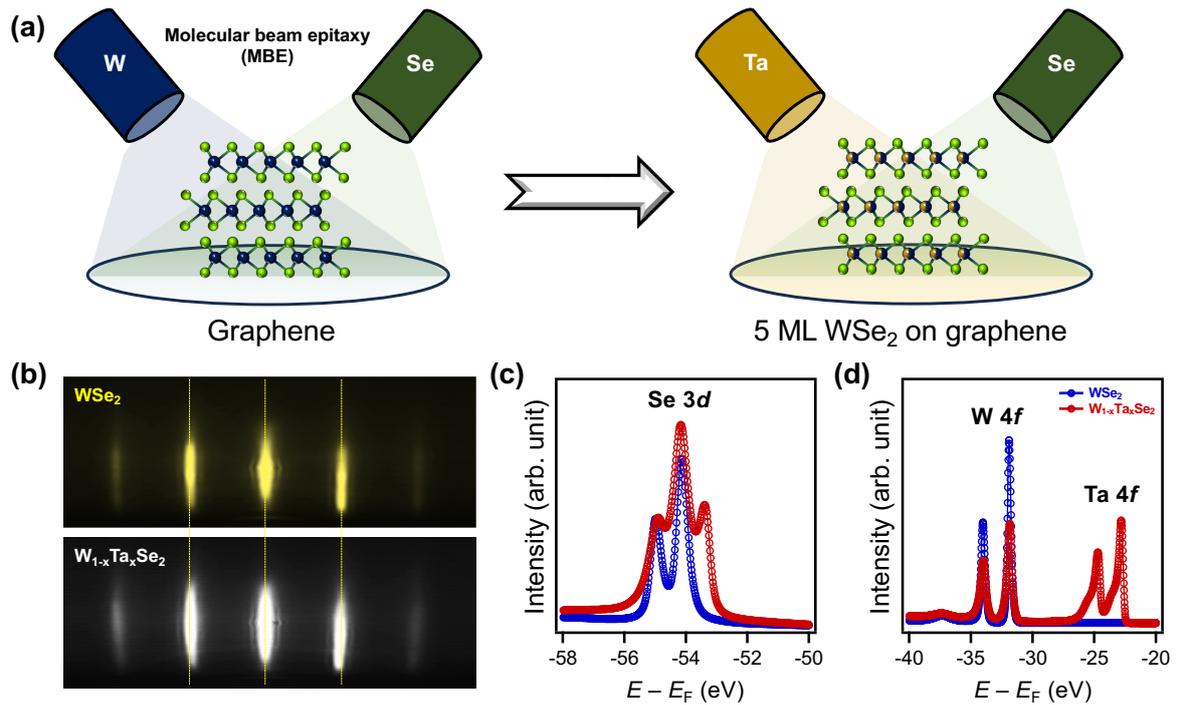

Figure 1. (a) Schematics for the MBE growth of 2$H$-WSe$_2$ (left) and 2$H$-W$_{(1-x)}$Ta$_{(x)}$Se$_2$ films (right) on a graphene substrate. (b) RHEED images of MBE-grown 2$H$-WSe$_2$ (top) and 2$H$-W$_{(1-x)}$Ta$_{(x)}$Se$_2$ films (bottom). Yellow dashed lines indicate the RHEED peak positions of 2$H$-WSe$_2$. (c-d) Core-level photoemission spectra from Se 3$d$- (panel c) and W and Ta 4$f$-levels (panel d) of 2$H$-WSe$_2$ (blue symbols) and 2$H$-W$_{(1-x)}$Ta$_{(x)}$Se$_2$ (red symbols) films taken using $s$-polarized 90 eV photons at RT.

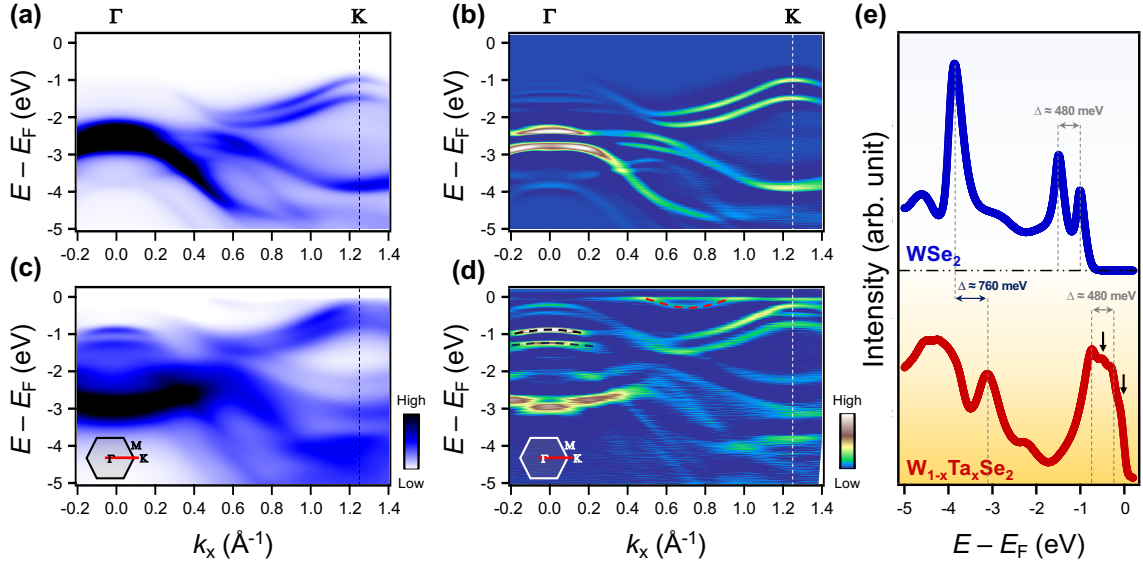

Figure 2. (a-b) An ARPES intensity map for 5 layers of MBE-grown 2$H$-WSe$_2$ (panel a) and its second derivative (panel b) taken along the Γ-K direction using $s$-polarized 70 eV photons at RT. (c-d) An ARPES intensity map of MBE-grown 2$H$-W$_{(1-x)}$Ta$_{(x)}$Se$_2$ films (panel c) and its second derivative (panel d) taken along the Γ-K direction using $s$-polarized 70 eV photons at RT. Red and black dashed guidelines present the electron and hole bands of ML 1$H$-TaSe$_2$, respectively [16,17]. The flat feature near $E_F$ is an artificial effect induced by the second-derivative processing. (e) Energy distribution curves (EDCs) taken at the K point from ARPES data of 2$H$-WSe$_2$ (blue curve) and 2$H$-W$_{(1-x)}$Ta$_{(x)}$Se$_2$ (red curve). The position and splitting of the peaks were determined by Lorentzian fitting of the ARPES spectra.

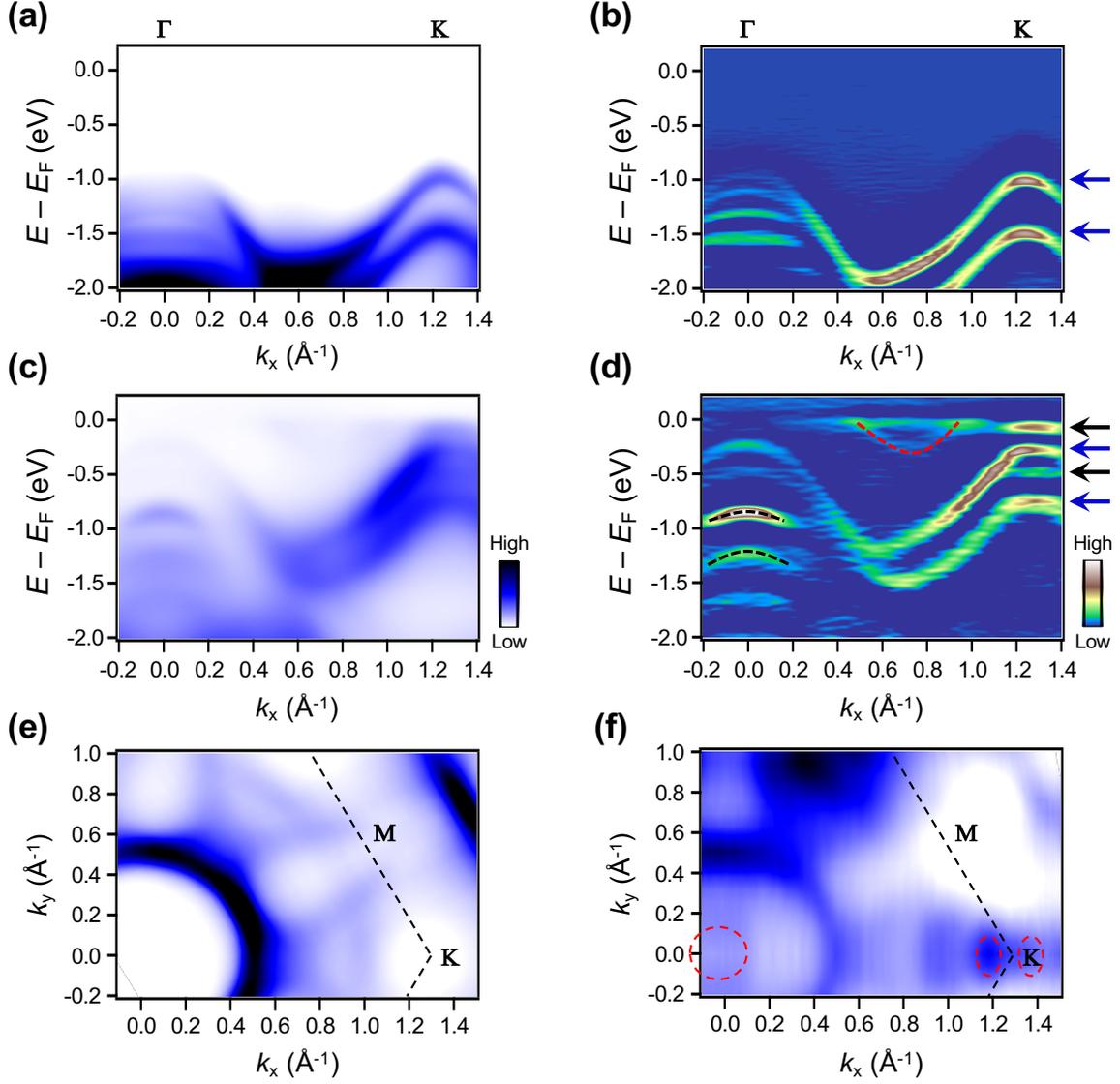

Figure 3. (a-d) Zoomed-in ARPES intensity maps of (panel a) 5 layers of MBE-grown 2$H$-WSe$_2$ and 2$H$-W$_{(1-x)}$Ta$_{(x)}$Se$_2$ films (panel c) taken along the Γ-K direction using $s$-polarized 70 eV photons at RT. Panels b and d show their second derivative intensity maps. Blue and black arrows indicate the split bands with strong and weak spectral intensity at the K point, respectively. Red and black dashed guidelines are the band structures from ML 1$H$-TaSe$_2$. (e-f) Fermi surfaces of MBE-grown ML 1$H$-TaSe$_2$ on bilayer graphene substrate (panel e) and 2$H$-W$_{(1-x)}$Ta$_{(x)}$Se$_2$ films (panel f). Red dashed ovals indicate the Fermi surface formed by 2$H$-W$_{(1-x)}$Ta$_{(x)}$Se$_2$ bands.